\documentstyle[12pt,aasms4]{article}
\def\Q{{\bf Q}}
\def\lab#1{\label{#1}}
\def\br{{\bf r}}
\def\be{\begin{equation}}
\def\ee{\end{equation}}

\def\bfo{\omega\hskip-7.5pt\omega
\hskip-7.6pt\omega
\hskip-7.7pt\lower.25pt\hbox{$\omega$}}
\def\bxi{\xi
\hskip-6pt
 {\xi}
\hskip-5.75pt\xi
\hskip-5.5pt\xi
\hskip-5.6pt\xi}
\def\la{\lambda}
\def\hz{{\hat z}}
\def\f{{\bf f}}
\def\gsim{\raise2.90pt\hbox{$\scriptstyle
>$} \hspace{-6.4pt}
\lower.5pt\hbox{$\scriptscriptstyle
\sim$}\; }
\def\lsim{\raise2.90pt\hbox{$\scriptstyle
<$} \hspace{-6pt}\lower.5pt\hbox{$\scriptscriptstyle\sim$}\; }
\def\beq{\begin{eqnarray}}
\def\tfrac#1#2{{\textstyle{#1\over #2}}}
\def\z{{\bf z}}
\def\obet{{\beta}}
\def\k{{\bf  k}}
\def\hbz{\hat {\bf z}}
\def\hbw{\hat {\bf w}}
\def\Z{{\bf Z}}
\def\fbf{{\bf f}}
\def\lra#1{\left\langle #1\right\rangle}
\def\bj{{\bf j}}
\def\lrp#1{\left( #1\right)}

\def\eeq{\end{eqnarray}}
\def\ep{\epsilon}
\def\part{\partial}

\def\w{{\bf w}}
\def\nn{\nonumber}

\def\x{{\bf x}}
\def\l{\left}

\def\etal{{\it et al.\ }}

\def\r{\right}

\def\v{{\bf v}}
\def\k{{\bf k}}
\def\b{{\bf b}}
\def\B{{\bf B}}
\def\V{{\bf V}}
\def\OV{\overline{\V}}
\def\OB{\overline{\B}}
\def\ob{\overline{B}}
\def\ao{\alpha\hbox{-}\Omega}

\def\half{\tfrac12}
\def\third{\tfrac13}
\def\gammak{\gamma_k}

\def\w{{\bf w}}
\begin{document}
\begin{center}
{\Large\bf Nonlinear $\alpha$-Effect in Dynamo Theory}
\bigbreak
{\large\bf by}
\medbreak
{\large\bf George B. Field, Eric G. Blackman, and Hongsong Chou}
\end{center}
\begin{abstract}
The standard two-scale theory of the dynamo coefficient  $\alpha$ in
incompressible isotropic helical MHD turbulence is extended to include
nonlinear effects of $\OB$, the large-scale magnetic field.  We express
$\alpha$ in terms of statistical quantities that can be calculated from
numerical simulations of the case
$\OB=0$.  For large magnetic Reynolds numbers our formula agrees
approximately with that of Kraichnan (1979), but disagrees with that of
Cattaneo and Hughes (1996).
\end{abstract}

\section{Introduction}
Magnetic fields of galaxies are important in astrophysics and
cosmology.  In astrophysics, because they enable fast particles to be
accelerated and trapped, and affect the dynamics of star formation.  In
cosmology, because if   galactic magnetic fields do not
originate in the modern era, they could be relics from the early universe,
carrying information about that period.

Since Parker's (1955) paper on the $\ao $  dynamo, and his application of
dynamo theory to the Galaxy (1971), most workers have attributed the
origin of the magnetic fields of disk galaxies to the operation of an $\ao$
turbulent dynamo (Ruzmaikin, Shukurov \& Sokoloff 1988).
However, the standard theory of the dynamo  has always been open to 
criticism.  For example, Piddington (1970, 1972abc, 1975ab) argued that the
small-scale magnetic field produced by the small-scale turbulence required
by the theory would rapidly grow to equipartition, quenching dynamo action,
and this point has been demonstrated numerically by Kulsrud and Anderson
(1992).  

Recently a series of papers 
(Vainshtein \& Rosner 1991; Cattaneo \& Vainshtein 1991; Vainshtein and
Cattaneo 1992; Tao, Cattaneo \& Vainshtein 1993; Cattaneo 1994; and
Cattaneo \& Hughes 1996; see also Seehafer 1994, 1995) have argued that
dynamo action is quenched if the large-scale magnetic field  $\ob$ in
velocity units exceeds  a critical value,
$\ob_c =
 R^{-1/2}_Mv_0$, where $R_M=v_0L/\la$ is the magnetic Reynolds number
of the turbulence,
$v_0$ is the turbulent velocity at the outer scale  of the turbulence $L$, and
$\la=\eta c^2/4\pi$ is the magnetic diffusivity, with $\eta$ the
resistivity.  They argue that this result is supported by  direct  numerical
simulations of MHD  incompressible turbulence (Cattaneo \&
Hughes 1996).  If they are correct, the classical $\alpha$-$\Omega$ dynamo
theory based upon Parker's (1955) paper and developed by the Potsdam group
(Krause and R\"adler 1980; see also Moffatt 1978 and Parker 1979), which
applies to weak large-scale magnetic fields, is not applicable to
present-day galaxies, as
  $\ob$ is observed to greatly exceed $R^{-1/2}_M v_0$.   

The argument by Cattaneo and his collaborators depends upon their analysis
of nonlinear interactions which are surely present.  The effects of such
interactions have also been considered independently in a series of papers
on MHD turbulence by Pouquet and her collaborators (Frisch, Pouquet,
L\'eorat, \& Mazure 1975; Pouquet, Frisch,
\& L\'eorat 1976; Pouquet \& Patterson 1978; Meneguzzi, Frisch \& Pouquet
1981)  using spectral methods.  By and large
this work supports the applicability of the classical theory even for
large-scale fields
$\ob$ approaching $v_0$, which encompass those observed.  Chandran (1996)
confirms the results of Pouquet, Frisch,
\& L\'eorat (1976) using  a different spectral method. 

Since $R_M\gg 1$ in most astrophysical situations (as large as $10^{20}$ in
the interstellar medium of our Galaxy) the difference between the
critical field  advocated by Cattaneo and his collaborators, $\ob_c\sim
R^{-1/2}_Mv_0$,  and that implied by Pouquet and her collaborators,
$\ob_c\sim v_0$, is crucial.

In this paper we extend the  calculation of one of the dynamo coefficients,
$\alpha$, of the classical theory to include arbitrary values of $\ob$.  Our
result contains two terms, one which resembles that advocated by
Cattaneo and his collaborators in its dependence upon  
$R_M$.  The other
resembles a formula proposed  by Kraichnan (1979) on the basis of a simple
model incorporating damping by nonlinear interactions.  For
$R_M\gg 1$, the latter term dominates, so we find that the classical
result for $\alpha$ applies for
$\ob$ of the order of  $v_0$.
We  plan to extend the 
simulations of Cattaneo and his collaborators to other values of
$R_M$ in order to clarify our disagreement with them.

We confine discussion to
the alpha effect, and omit any discussion of
turbulent diffusion, which is also controversial.

Before proceeding to the nonlinear theory we  review the classical
linear theory.  The classical theory is based on a clear separation of scales
between the scale $L$ of the dominant turbulent motions (here called  the
``outer scale"), and the size
$D$ of the system, with $L\ll D$.  The large-scale magnetic field   $\OB$ 
satisfies an induction equation,  
\be
\part_t \OB = \nabla\times (\OV\times \OB)+ \nabla\times \lra{\v\times
\b} + \lambda \nabla^2 \b\; , \ee
where $\OV$ is the large-scale velocity field (differential rotation in the
case of a  galaxy), and $\lra{\v\times \b}$, called the ``turbulent emf,'' is the
spatial average of the cross product of the small-scale velocity $\v$ and
the small-scale magnetic field $\b$ over a scale much smaller than $D$,
but much larger than $L$.

The first term in (1) leads to the so-called ``$\Omega$-effect,'' according
to which lines of force of $\OB$ are stretched by the differential rotation,
creating a growing toroidal field from a poloidal one.  This term is not
controversial, and will not be discussed further here. Controversy 
centers on the turbulent emf,
$\lra{\v\times \b}$, which enables a growing toroidal field to feed back
into the poloidal direction,   giving exponential
amplification of the large-scale $\OB$.

The evaluation of  $\lra{\v\times \b}$ is usually
restricted to incompressible isotropic turbulence.  On might assume that if 
the turbulent velocity field $\v$ is isotropic, the small-scale
magnetic field $\b$ would be isotropic also,  so that the $c$-component of
the emf would be
\be
\lra{\v\times \b}_c = \ep_{cde} \lra{v_d b_e} = \tfrac13
\ep_{cde}\delta_{de}\lra{\v\cdot\b} = 0\; , \ee
in the light of an identity for isotropic tensors (Krause \&
R\"adler 1980).
However, this is not correct, because $\OB$,   being
anisotropic, induces a term in $\b$ that is not isotropic even though
$\v$ is isotropic.  Evaluating this term requires   the induction
equation for $\b$, which can be found by writing the   induction equation
for  the total magnetic field $\B=\OB+\b$ and separating off the
small-scale parts to give
\be
\part_t \b = \nabla\times (\v\times \OB) + \nabla\times (\OV\times \b) +
\nabla\times (\v\times \b)
-\nabla \times \lra{\v\times \b}+\lambda \nabla^2 \b\; . \ee

  The second term on the right in (3)
represents a change of the reference frame to that moving at the mean
velocity $\OV$, in which
$\v$ may be assumed to be isotropic; we assume $\V=const.$, so we make
that change and henceforth omit the term.  The fourth term on the right is a
large-scale quantity, so its time integral, being large scale, will contribute
nothing when crossed wtih $\v$ (small-scale) and averaged, and so is
neglected henceforth. The first term   is
$-\v\cdot\nabla\OB +\OB\cdot \nabla\v$.  As the $\v\cdot\nabla\OB$ term
ultimately leads to turbulent diffusion of $\OB$, which is not our main
interest here, we omit it, leaving
\be
\part_t \b = \OB\cdot \nabla\v +\nabla\times (\v\times \b)+\lambda
\nabla^2 \b\;. \ee 

The second term in (4) is neglected in the classical discussions in a step
referred to as the first order smoothing approximation, or FOSA.  Krause \&
R\"adler (1980) showed that this is legitimate if $\lambda$ is large, so
$R_M$ is small, because then Ohmic diffusion keeps $b$ small, and the first
term dominates the second.  However, because $R_M$ is large in
astrophysics, this case is not relevant here.  They also discuss the case
when the Strouhal number
\be
S = {v_0\tau\over L} \label{new5}
\ee
is small, where $\tau$ is the correlation time for eddies at the outer scale
of the turbulence.  Because $L/v_0=t_{ed}$, the eddy turnover time,
$S=\tau/t_{ed}$, and one might suppose that $S=O(1)$.  In fact, it is
observed experimentally that $S\simeq 0.2-0.3$ in ordinary hydrodynamic
turbulence (Pope 1994).  Although this value is not as small as one would
like, it provides a way to approximate (4).  To understand how, it is
important to distinguish between $\b^{(0)}$, the small-scale magnetic field
when $\OB=0$, and $\b^{(1)}$, the perturbation to $\b$ when a small $\OB$
is present.  If the turbulent velocity field $\v$ is isotropic, $\b^{(0)}$ is,
and as shown in (2), its contribution to $\lra{\v\times \b}$ vanishes. 
However, $\b^{(1)}$ is not isotropic, and its contribution to $\lra{\v\times
\b}$ does not vanish.

If we write $\b=\b^{(0)}+\b^{(1)}$, $\b^{(0)}$ is governed by
\be
\part_t \b^{(0)}=\nabla\times \lrp{\v^{(0)}\times \b^{(0)}}
+\lambda\nabla^2\b^{(0)}\;,  \label{5}
\ee
where $\v^{(0)}$ is the isotropic turbulent velocity.  Parker (1979; p.~511)
shows that if $\lambda$ is small (the astrophysical case), (\ref{5}) leads
to exponential growth of $b^{(0)}$ with a time constant $\tau/S^2$
 until (in velocity
units) it begins to approach $v^{(0)}$.  At that point, we expect that back
reaction due to the Lorentz force associated wtih the large value of the
small-scale field $b^{(0)}$ will result  in a steady state in which the energy
driving the turbulence at the outer scale (buoyancy forces in stars,
supernova explosions  in galaxies) is balanced by a nonlinear cascade  to
smaller scales, where it is dissipated by viscosity and/or Joule heating.  
This expectation is confirmed by the Pouquet \etal (1976) and by direct
simulations, {\it e.g.}\ Cattaneo and Vainshtein (1991), Cattaneo (1994),
and Cattaneo and Hughes (1996).
 As
we explain later, we will take this steady state as the base state which is
perturbed by
$\OB$.  The first term in (\ref{5}) mediates exchange of magnetic energy
with kinetic energy in MHD turbulence, as discussed in Appendix A, and it
certainly cannot    be neglected.  

However, we now argue that the corresponding term in the equation for
$\b^{(1)}$ {\it can} be neglected, as follows.  Evidently $\b^{(1)}$ is
governed by the parts of (4) which are of first order in $\OB$, namely,
\be
\part_t \b^{(1)} = \OB\cdot \nabla\v^{(0)}+\nabla\times \lrp{\v^{(0)}\times
\b^{(1)}}+\lambda \nabla^2 \b^{(1)}\; . \label{6}
\ee
In the rest of this section, we assume that $R_M\gg 1$,
so that we may neglect the third term.  We formally integrate (\ref{6}) to
get
\beq
\b^{(1)} (\x,t)&=& \OB\cdot \nabla\int^t_{-\infty}dt_1 \v^{(0)}(\x,t_1)\nn\\
&&+ \int^t_{-\infty}dt_1\nabla\times \l[\v^{(0)}(\x,t_1)\times
\b^{(1)}(\x,t_1)\r]\; .  \label{7a}
\eeq
This does not seem useful, because the desired quantity, $\b^{(1)}$, appears
under the integral.  However, we can show that the integral is much smaller
than $\b^{(1)}$ itself, which appears on the left-hand side, if $S\ll 1$.

Since  
$\v^{(0)}$ is a stochastic function of $t$, only those 
parts of the integral in (\ref{7a}) which come from
times $t_1$ which
differ from $t$ by less than a correlation time $\tau$ will correlate
significantly with $\v^{(0)}$ in the turbulent emf, $\lra{\v^{(0)}(\x,t)\times
\b^{(1)}(\x,t)}$.  Hence we can replace the lower limit on the integral by
$t-\tau$, and then estimate the integral by
$\tau$ times the integrand at $t_1=t$.  Since the order of magnitude of
$\nabla$ is $L^{-1}$, the order of magnitude of the second term in (\ref{7a})
is
\be
{\tau v^{(0)}\over L}b^{(1)} (t)=Sb^{(1)} (t)\; , \label{7b}
\ee
whose ratio to the magnitude of the left hand side of (\ref{7a}) is $S$.  If,
as we shall argue later, $S<1$, we can neglect the second term in (\ref{7a}),
and hence in (\ref{6}), so, still neglecting diffusion, (\ref{6}) becomes
\be
\partial_t\b^{(1)}=\ob \cdot \nabla\v^{(0)}\; . \label{7c} 
\ee
Effectively $b^{(1)}\sim S\ob$, so if $S<1$, $b^{(1)}<\ob$, even when
$\ob\ll v^{(0)}$.  However this says nothing about $b^{(0)}$, which as we
have stressed, approaches $v^{(0)}$ in value, even if $v^{(0)}\gg \ob$. 
Hopefully this discussion clarifies a point which has led to confusion in the
past. 

 The solution to (\ref{7c}) can be written
in  component form as 
\be
b^{(1)}_e (t) = \ob_p \part_p \int^t_{-\infty} dt' v_e^{(0)} (t')\; , \lab{9}\ee
so that 
\be
\lra{\v^{(0)}\times \b^{(1)}}_c = \ep_{cde} \lra{v^{(0)}_d (t) \ob_p \part_p
\int^t_{-\infty} dt' v_e^{(0)} (t')} \;.\lab{10}\ee
Averaging commutes with integration and differentiation,   spatial
differentiation commutes with time integration, and $\ob$ is independent
of time on the scale $\tau$, so
\be
\lra{\nabla\times \b}_c = \ob_p \ep_{cde} \int^t_{-\infty} dt' \lra{v^{(0)}_d
(t)\part_p v^{(0)}_e(t')}\; . \lab{11}\ee
Since $\v^{(0)}$ correlates with its derivatives for a time of order $\tau$,
the integral is of order $\tau$ times the average taken at a given time 
(Krause \& R\"adler 1980).  To make further progress in what follows, we
need an explicit dependence of the indicated correlation on $t$ and $t'$. 
Because the turbulence is assumed to be steady, it can depend only on
$t-t'$, so a convenient representation is 
\be
\lra{v^{(0)}_d (t)\part_p v_e^{(0)}(t')} = \lra{v^{(0)}_d \part_p v^{(0)}_e}
e^{-|t-t'|/\tau}\; ,\lab{12}\ee
where the common time argument in the second average has been omitted
because the turbulence is presumed steady.  Hence $\lra{\v\times \b}$ is
given  by 
\be
\lra{\v\times \b}_c = \ob_p \tau \ep_{cde} \lra{v^{(0)}_d \part_p
v^{(0)}_e}\; . \lab{13}\ee
At this point we use the assumption that  the
velocity
$\v^{(0)}$ is distributed isotropically.  According to Krause
\& R\"adler (1980), a third-rank isotropic tensor like that in (\ref{13}) can
be written
\be 
\lra{v_d^{(0)} \part_p v^{(0)}_e} = \tfrac16 \ep_{dpe} \lra{\v^{(0)} \cdot
\nabla\times \v^{(0)}}\; ,
\lab{14}
\ee
 so from (\ref{11}),
\be
\lra{\v\times \b} = \alpha \OB\; ,\lab{15}
\ee
where
\be
\alpha = -\tfrac{1}{3} \tau \lra{\v^{(0)} \cdot \nabla\times \v^{(0)}}
\lab{16}\ee
is the classical expression for the dynamo coefficient (Krause \& R\"adler
1980, Moffatt 1978), but with the additional feature that the turbulent
velocities indicated refer to the zero-order state.  The quantity in angular
brackets, a pseudoscalar, is known as the kinetic helicity of the turbulent
flow $\v^{(0)}$.

In an independent development,
Pouquet, Frisch \& L\'eorat (1976) calculated magnetic energy
spectra for MHD turbulence, solving the spectral equations using a
closure method known as the  EDQNM (Eddy-Damped Quasi-Normal Markovian)
approximation.   They found that if 
$\lra{\v^{(0)}\cdot
\nabla\times \v^{(0)}}$ vanishes, the magnetic energy spectrum $E^M_k$
peaks near $k_0$, the wave number at which turbulent energy is
injected, and reaches a steady state in which the total energy $E_k$, the
sum of the  kinetic energy $E^V_k$ and the magnetic energy
$E^M_k$, cascades  to higher wave numbers, ultimately to be lost to Ohmic
dissipation and/or viscosity.

If, on the other hand, $\lra{\v^{(0)}\cdot \nabla\times \v^{(0)}}\ne 0$, they
found that $E^M_k$ inverse cascades, accumulating at an ever-decreasing
wave number.  They attributed this  to the  turbulent dynamo
effect described above,  operating in the nonlinear regime. 
In the special case that the wave number $k$ of interest is $\ll k_0$, so
that there is a clear separation of scales, they find an approximate
expression for a quantity $\alpha_k$ governing the growth of
$E^M_k$, where
\be
\alpha_k = -\tfrac43 \int^\infty_{k/a} dq \theta_{kqq} (H^V_q - q^2
H^M_q)\;. \lab{17}\ee
 Here $a$ is a small parameter, $\theta_{kqq}$ is
an effective correlation time for modes of wave number $q$, and $H^V_q$ and
$q^2 H^M_q$ are the spectra corresponding to the kinetic helicity correlation
function
$\lra{\v^{(0)} (\x) \cdot \nabla\times \v^{(0)}(\x+\bxi)}$ and the current
(Keinigs 1983)  helicity correlation function $\lra{\b^{(0)}(\x)\cdot
\nabla\times
\b^{(0)}(\x+\bxi)}$, respectively.  (Here $\b$ is in velocity units, obtained by
dividing $\b$ by
$\sqrt{4\pi\rho}$, so that $\bf b$ is the vector Alfv\'en velocity.)  
The term adopted by  Pouquet \etal for  (\ref{17}),  
the {\it residual torsality},  has not reappeared in the
literature.  Here we note that the first term in (\ref{17}) is similar to
(\ref{16}), so that classical dynamo theory can be interpreted as an  inverse
cascade in helical turbulence.  The second term does not appear in the
classical result, but Montgomery and Chen (1984) verified it using
calculations in real rather than $k$ space.  
Pouquet
\etal thus found  that dynamo action takes place in the fully nonlinear regime,
with no restriction as to the magnitude of
$\b$ or $\OB$, and that the classical expression for
$\alpha$ must be modified by the addition of the second term in (\ref{17}).

\section{The Method of This Paper}
In this paper we extend the classical analysis of the $\alpha$ effect into the
regime of large $\ob$.  We use the standard two-scale approximation in real
space, and at several points we refer to the results obtained by Pouquet
\etal (1976) using the spectral closure method.  We find that when a clear
distinction is made between the quantities $\v$ and $\b$ on the one hand,
and their values when $\OB=0$, $\v^{(0)}$ and $\b^{(0)}$, nonlinearity due to
$\ob$ appears in a straightforward manner.  A key feature of our derivation
is the assumption that  correlations are damped by nonlinear
interactions, following Pouquet \etal (1976) and  Kraichnan (1979).
As shown by Kichanitov (1985), such an assumption can be justified by
application of renormalization group methods.

Our results depend upon the damping rate at each wave number $k$,
$\gammak$, the spatial spectra of
$\v^{(0)}$ and
$\b^{(0)}$, the value of $\ob$, and the value of $\la$, the magnetic
diffusivity, expressed in terms of the magnetic Reynolds number $R_M$,
which for convenience we
assume  to be equal to the Reynolds number $R$.  We find that $\alpha$ is a
well-behaved function of the parameters, and that the behavior for larger
$R_M$ is similar to that predicted by Kraichnan (1979).

We take as our base state  fully-developed MHD turbulence
driven by external forces and in a steady state as a result of a turbulent
cascade to large wave numbers, but with $\ob=0$.  As demonstrated by
Pouquet
\etal (1976), in  such turbulence there is  approximate
equipartition between the magnetic energy $E^M_k$ and the kinetic energy
$E^V_k$ for wave numbers $k\gsim 3k_0$, where $ L= k_0^{-1}$ is the outer
scale of the turbulence.  As shown by Parker (1979, p.~513), approach to the
steady state occurs  on the scale of the eddy turnover time
$t_{ed} = L/v^{(0)}$, which, as stressed by Kulsrud
\& Anderson (1992), is much shorter than the dynamo growth time. 
According to the calculations of Pouquet {\it et al.}, the
saturation which occurs at small scales (large $k$) does not prevent the
increase of magnetic energy on scales larger  than $L$ if the turbulence is
helical.  It is important to note that the growth of a large-scale field as a
consequence of an
$\alpha$ effect does not substantially modify the spectra of
$E^M_k$ calculated by Pouquet \etal (1976) for $k$ of the order of $k_0$
(their Fig.~8), so our concept of a base state independent of $\OB$ is valid. 
We assume that the properties of the base state  can be calculated once and
for all by  numerical simulations. Our results then allow us    to
calculate
$\alpha$ in terms of those properties.

Following Montgomery and Chen (1984), we  present our calculations in
terms of the Elss\"aser variables $\z^\pm = \v\pm \b$, (Biskamp 1993) and
have checked our results by carrying out the calculation  in terms of $\v$
and
$\b$.  Elss\"aser variables  are
naturally adapted to the problem, shortening the calculation substantially. 
More important, we show in Appendix A that unlike the kinetic energy
$E^V_k$ associated with $\v$ and the magnetic energy $E^M_k$ associated
with $\b$, both $E^+_k$ and $E^-_k$   cascade directly in isotropic
turbulence, allowing us to employ a single  damping constant $\gamma_k =
\gamma_k^+=\gamma_k^-$ to describe the effects of the nonlinear terms in
the base state.

Our goal is to calculate the turbulent emf, $\lra{\v\times \b}$.  Since
\be
\v=\tfrac12 (\z^++\z^-)\lab{18}
\ee
and
\be
\b=\tfrac12 (\z^+-\z^-)\; ,\lab{19} \ee
we have
\beq
\lra{\v\times \b}_c &=& \tfrac14 \lra{\lrp{\z^+ + \z^-} \times
\lrp{\z^+-\z^-}}_c = -\tfrac12 \lra{\z^+ \times \z^-}_c\nn\\
&=& -\tfrac12 \lra{\ep_{cde} z^+_d z^-_e}\; . \lab{20}\eeq
We use this formula in what follows.

\section{Evolution of the Turbulent Fields}
As explained above, the  correlation indicated in (\ref{20})  vanishes for
isotropic turbulence, but is nonzero when one takes into account the
perturbations of
$\z^+$ and $\z^-$ which are caused by $\OB$.  To obtain these, we  
consider the dynamical equations for $\z^\pm$, or equivalently, $\v$ and
$\b$.

When Ohmic dissipation is included, the induction equation for $\b$ is, from
(3),
\be
\part_t \b=-\v\cdot
\nabla\OB +\OB \cdot \nabla\v-\OV\cdot \nabla\b+\b\cdot
\nabla\OV-\v\cdot \nabla \b+\b \cdot \nabla\v + \la \nabla^2 \b\; ,\lab{21}
\ee where $\lambda$ is the magnetic diffusivity. As explained in \S 1,
adopting in a  frame of reference   moving with
$\OV$ eliminates the third term on the right, and the
fourth term can be neglected with respect to the sixth because we
assume that the size of the system $S\gg L$.  The remaining terms can be
written in terms of
$\B=
\OB +\b$ as
\be
\part_t \b = -\v \cdot \nabla\B + \B\cdot \nabla\v +\la \nabla^2\b\;
.\lab{22}\ee

The classical theory ignores the effect of $\ob$ on the velocity, on the
grounds that the Lorentz force associated with $\ob$ is of order $\ob^2$,
hence negligible in the limit $\ob\to 0$. However, as we have explained
above, $b^{(0)}$ grows quickly to approximate equipartition, so even in a
first-order calculation, a Lorentz force proportional to $\ob b^{(0)}$ must
be included.  It is therefore essential to consider the effect of $\ob$ on
$\v$; we shall do so to all orders in $\ob$.  To do this, we use the equation
of motion for the small-scale velocity $\v$:
\be
\part_t \v = -\v \cdot \nabla\v - \nabla p +\B\cdot \nabla\B -
\nabla\tfrac12 B^2 +\nu \nabla^2 \v +\fbf\;,
\lab{23}\ee
where $\fbf$ is the applied force per unit mass and $\nu$ is the
kinematic viscosity;  $\B$, the magnetic field $\div \sqrt{4\pi
\rho}$, is in velocity units. Hence
\be
\part_t \v = -\v\cdot \nabla\v - \nabla P +\B\cdot \nabla\B + \nu \nabla^2
\v+\fbf\;, \lab{24}\ee
where
\be
P = p+\tfrac12 B^2\; .\lab{25}\ee
We define the Elss\"aser variables for the field $\B$ as
\be
\Z^\pm = \v\pm \B = \v\pm \b\pm \OB =\z^\pm \pm \OB\; ,\lab{26} \ee
so that
\be
\v=\tfrac12 (\Z^++\Z^-)\lab{27}\ee
and
\be
\B= \b+\OB=\tfrac12 (\Z^+-\Z^-)\; . \lab{28}\ee
Then (\ref{22}) becomes
\beq
\tfrac12 \part_t \Z^+ - \tfrac12 \part_t \Z^- &=& - \tfrac14 (\Z^++\Z^-)
\cdot \nabla (\Z^+-\Z^-) \nn\\
&&\; + \tfrac14 (\Z^+-\Z^-)\cdot \nabla (\Z^++\Z^-) \nn\\
&& \; + \tfrac12 \la \nabla^2 (\Z^+-\Z^-) \; . \lab{29}\eeq
(Note that we have set $\part_t \OB=0$ because $\OB$ varies only on the
long time scale.)  Equation (\ref{24}) becomes
\beq
\tfrac12 \part_t \Z^+ +\tfrac12 \part_t \Z^- &=& -\tfrac14 (\Z^++\Z^-)\cdot
\nabla(\Z^++\Z^-)\nn\\
&&\; + \tfrac14 (\Z^+-\Z^-) \cdot \nabla(\Z^+-\Z^-) \nn\\
&&\; + \tfrac12 \nu \nabla^2 (\Z^++\Z^-) - \nabla P+\fbf\; . \lab{30}\eeq
Adding and subtracting (\ref{29}) and (\ref{30}) yields
\be
\part_t \Z^\pm = -\Z^\mp \cdot\nabla\Z^\pm + \tfrac12(\nu+\la) 
\nabla^2 \Z^\pm +\tfrac12 (\nu-\la) \nabla^2 \Z^\mp - \nabla P +\fbf\;. 
\lab{31}\ee
The dissipative coupling between $\Z^+$ and $\Z^-$ vanishes
in the special case  that the magnetic Prandtl number $\lambda/\nu$
equals unity.  For simplicity  we assume that is the case in what follows;
we note that  both Pouquet
\etal (1976) and Cattaneo and Hughes (1996) also made the same assumption.
Chou and Fish (1998) discuss the case $\lambda/\nu\ne 1$.

From (\ref{26}) the nonlinear term in (\ref{31}) is
\beq
\Z^\mp \cdot \nabla \Z^\pm &=& (\z^\mp \mp \OB)\cdot \nabla(\z^\pm \pm
\OB)\nn\\
&=& \z^\mp \cdot \nabla\z^\pm \mp \OB\cdot \nabla\z^\pm\; , 
\lab{32}\eeq
where we have neglected $\nabla\OB$ for the reasons given previously. 
Hence (\ref{31}) becomes
\be 
\part_t \z^\pm =  - \z^\mp \cdot \nabla \z^\pm +\la \nabla^2 \z^\pm -
\nabla P +\fbf \pm \OB\cdot \nabla \z^\pm\; , \lab{33}\ee
where   we have neglected $\part_t\OB$ for the reason given
previously.

We adopt a perturbation expansion in $\OB$, in which the zero-order
variables $\z^{\pm(0)}$ describe the turbulence exactly if $\OB$ is zero. 
Therefore $\z^{\pm(0)}$ satisfies
\be
\part_t \z^{\pm(0)} = -\z^{\mp(0)}\cdot \nabla\z^{\pm(0)}+\la \nabla^2 
\z^{\pm(0)} - \nabla P^{(0)} +\fbf \;, \lab{34}\ee
where
\be
P^{(0)} = p^{(0)} + \tfrac12 B^2\big\vert_{\OB=0} = p^{(0)} +\tfrac12
(b^{(0)})^2\; .\lab{35}\ee
If we assume that the driving force is independent of
$\OB$,   $\fbf$ has the same value in both the zero-order and the
perturbed state.

 Equation (\ref{34}) can be solved numerically for a variety of
initial conditions, and so in principle any averages of zero-order quantities
required can be computed.  Our task is then to compute $\lra{\v\times \b}$,
which depends on the perturbations induced by $\OB$, in terms of averages
over zero-order quantities.

We let
\be
\z^\pm = \z^{\pm(0)} +\z'^\pm\; , \lab{36}\ee
where $\z'^\pm$ contains
perturbations of all orders in $\ob$.  Although in principle $\z'^\pm$ can  be
represented as a power series in $\ob$, we find that it is not necessary to
do so for the special case $\lambda/\nu=1$.  If we eliminate
$\fbf$ between (\ref{33}) and (\ref{34}), we find that
\beq
(\part_t - \la \nabla^2) \z'^\pm &=& - \z'^\mp \cdot
\nabla\z^{\pm(0)}-\z^{\mp(0)} \cdot \nabla \z'^\pm - \z'^\mp \cdot
\nabla\z'^\pm \nn\\
&&\; - \nabla (P-P^{(0)}) \pm \OB \cdot \nabla \z^\pm\; . \lab{37}\eeq

Inspecting (\ref{37}), we see that the left hand side and the term $\OB\cdot
\nabla \z'^\pm$ on the right are linear in $\z'^\pm$, hence easy to deal with. 
If
$R_M$ is moderately large, we can ignore the term
$\lambda\nabla^2\z'^\pm$ in  assessing  the order of magnitude of the
remaining terms.  Putting aside the pressure term for the moment, we can
write (\ref{37}) in the symbolic form
\be
\partial_t z' = L^{-1} \l[z'z^{(0)}, (z')^2, \ob z^{(0)},\ob z'\r]\;,\lab{38}\ee
where the commas separate terms of potentially different orders, and
where we have combined the first two terms on the right of (\ref{37}) into one. 
As in our discussion of the classical case, we take $\Delta z' = \int dt
 \partial_z z'$, the change in $z'$ after one correlation time $\tau$,  to
equal $z'$ in order of magnitude, by definition.  Then (\ref{38}) implies that
\be
z' = {\tau\over t_{ed}}\l[z', (z')^2/z^{(0)}, \ob,\ob z'/z^{(0)}\r]\;,\lab{39}\ee
where $t_{ed}\equiv L/z^{(0)}\simeq L/v^{(0)}$.

In the classical discussion it is assumed that because $\ob$ is small, back
reaction by the Lorentz force can be neglected, so the motions are
hydrodynamic in nature.  That allowed us to use the fact that $\tau/t_{ed}$
is small (see below). That is not really true, because, as we discussed
earlier, the zero-order state quickly approaches equipartition.
We want to argue, however, that even in the nonlinear case $\tau/t_{ed}$ is
a small number.  To do so, we appeal to the nonlinear calculations of
Pouquet \etal (1976) regarding the zero-order state.  They showed that the
magnetic energy spectrum $E^M_k$ is in equipartition with the kinetic
energy spectrum $E^V_k$ for $k\gsim 3k_0$, where $k_0$ is the wave
number at which kinetic energy is being injected (the outer scale of the
turbulence).  In this range, back reaction is a major effect, causing the
turbulence to become a field of interacting Alfv\'en waves.  However, these
large wave numbers need not concern us, because in such waves, the current
is perpendicular to the small-scale field, and the vorticity is perpendicular
to the small-scale velocity, so that the two helicities vanish.

Of crucial importance for us, Pouquet \etal (1976) found that in the range
$k_0\lsim k\lsim 3k_0$, $E^M_k<E^V_k$, so that the motions are largely
hydrodynamic in character, with only a modest back reaction of the
magnetic field.  This range is important because it contains
most of the energy.  Pope (1994) states that in a pure hydrodynamic
turbulence it is experimentally observed that
\be
{\tau\over t_{ed}} = 0.2-0.3 \;.\lab{40}\ee
We assume that (\ref{40}) applies to the 
energy-containing eddies in MHD turbulence driven like that of Pouquet
\etal  Although $\tau/t_{ed}$ is not a very small number, we  may take it
to be a small parameter
$\epsilon$ in analyzing (\ref{39}).  Thus, (\ref{39}) becomes 
\be
z' \sim \ep\l[ z',(z')^2/z^{(0)},\ob,\ob z'/z^{(0)}\r] \;.\lab{41}\ee
It seems reasonable to assume that the balance in (\ref{41}) is between $z'$
and the third term on the right, so
\be
z'\sim \ep \ob \;.\lab{42}\ee
We check this by evaluating the first two terms under that assumption.  We
see that the ratio of the first to the third terms, and the ratio of the
second to the fourth term, are both $\ep$, so the first and second terms can
be neglected as a first approximation.  Note, however, that the ratio of the
fourth to the third term is 
\be z'/z^{(0)} = \ep \ob/z^{(0)}\sim \ep \ob/v^{(0)}\;
,\lab{43}\ee
which cannot be neglected because, although $\ep$ is small, we want a
result which is valid to all orders in $\ob/v^{(0)}$.  Indeed, this is the
source of the nonlinearity in our calculation.

Note that although we have used the short correlation time approximation
to simplify our equations for $z'$, no assumption is made regarding the
ratio of $b$ to $\ob$.  This is a step forward, because the classical
discussion has been justly criticized for ignoring back reaction, which
amounts to assuming that $b$ is small.

We now 
 apply
the divergence operator
$\nabla\cdot$ to (\ref{37}) with the first three terms on the right dropped. 
Because
$\OB$ is constant and $\nabla\cdot$ commutes with $\OB\cdot \nabla$,
the fact that
\be
\nabla\cdot \z^\pm = \nabla\cdot (\v\pm \b)=0 \lab{44}\ee
for incompressible turbulence then implies that
\be
\nabla^2 (P-P^{(0)})=0\; , \lab{45}
\ee
which for a homogeneous system implies that $P-P^{(0)}= const.$, so that  
\be
\nabla(P-P^{(0)})=0\; .\lab{46}\ee
Hence (\ref{37}) becomes
\be
(\part_t - \la\nabla^2 \mp \OB\cdot \nabla)\z'^\pm = \pm \OB\cdot \nabla
\z^{\pm(0)}\;. \lab{47}\ee
 We must solve this equation in order to calculate the turbulent emf
according to (\ref{20}).

The operator $\OB\cdot \nabla$ is best handled by
introducing the spatial Fourier transform
\be
\hat{\z}'^\pm (\k,t) = (2\pi)^{-3} \int d\x e^{-i\k\cdot \x} \z'^\pm
(\x,t)\;,
\lab{48}
\ee so that  when written in terms of the components of $\z$,  (\ref{47})
becomes
\be
(\part_t +\la k^2 \mp i\k \cdot \OB) \hz'^\pm_d (\k,t) = \pm i\k \cdot \OB
\hz_d^{\pm(0)} (\k,t) \;,\lab{49}
\ee
to which the solution is
\be
\hz'^\pm_d (\k,t) = \pm i\k\cdot \OB \int^t_{-\infty}dt_1 \exp \l\{ - \l[
\mp i\k \cdot \OB+\la k^2\r](t-t_1)\r\} \hz^{\pm(0)}_d(\k,t_1) \; ,\lab{50}
\ee
 where we have put the lower limit equal to $-\infty$ because we will
find that the short correlation time makes values of  $t_1$   significantly
smaller than
$t$ irrelevant.  The nonlinear dependence on  $\OB$ is evident here.

From (\ref{20}) and (\ref{36}), the term of zero order in $z^\pm$ drops out
according to (2), leaving 
\beq
\lra{\v\times \b}_c &=& - \tfrac12 \lra{ \ep_{cde}z^{+(0)}_d (\x,t) z'^-_e
(\x,t)} - \tfrac12 \lra{\ep_{cde} z'^+_d (\x,t)z^{-(0)}_e (\x,t)} \nn\\
&& \;  - \tfrac12 \lra{\ep_{cde} z'^+_d (\x,t) z'^-_e (\x,t)} \;.\lab{51}
\eeq
Inverting the Fourier transforms, we see from (\ref{48}) that
\be
z^{\pm(0)}_d (\x,t) = \int d\k e^{i\k \cdot \x}\hz_d^{\pm(0)} (\k,t) \; ,
\lab{52} \ee
and, from (\ref{50})
\be
z'^\pm_e (\x,t) = \pm i\int d\k' e^{i\k'\cdot\x} \k'\cdot \OB \int^t_{-\infty}
dt_1
\exp \l\{ - \l[ \mp i\k'\cdot \OB+\la k'^2\r] (t-t_1)\r\} \hz^{\pm(0)}_e
(\k',t_1) \; . \lab{53}
\ee
Hence the first term in (\ref{51}) is
\be  \tfrac12 i\ep_{cde} \lra{\int d\k \int d \k' e^{i(\k+\k') \cdot \x}
\k'\cdot
\OB \int^t_{-\infty} dt_1 \exp \l\{ - [i\k'\cdot \OB +\la k'^2 ] (t-t_1)\r\}
\hz^{+(0)}_d (\k,t) \hz^{-(0)}_e (\k',t_1)} \;. \lab{54}
\ee 
Because  averaging
commutes with integration, this can be written
\be
\tfrac12 i\int d\k \int d\k' e^{i(\k+\k')\cdot \x} \k'\cdot
\OB
\int^t_{-\infty} dt_1 \exp \l\{ - [i\k'\cdot \OB+\la k'^2] (t-t_1)\r\}
\lra{\ep_{cde} \hz^{+(0)}_d (\k,t)\hz^{-(0)}_e (\k',t_1)} \; . 
\lab{55}
\ee
The indicated correlation decreases rapidly to zero as $\k-\k'$ and $t-t_1$
go to zero.  In Appendix B we show that the following representation is a
reasonable one:
\be
\lra{\ep_{cde} \hz^{+(0)}_d (\k,t) \hz^{-(0)}_e (\k',t_1)} = \ep_{cde} R_{de}
(-\k) \delta (\k+\k') e^{-\gamma_k |t-t_1|} \;.
\lab{56}\ee
  In principle, $\gammak$ could be
different for $\z^+$ and $\z^-$ modes, but we show in Appendix A that in
turbulence in which the cross helicity $K={1\over 2} \lra{\v\cdot \B}$, an
ideal invariant, vanishes, the associated energies $E^+$ and $E^-$ cascade
directly and are equal.  It is therefore reasonable to assume that if $K=0$,
$\gammak^+=\gammak^-$.  Following Pouquet \etal we assume that $K=0$,
and thus, that $\gammak^+=\gammak^-$.   Note that because $\gammak$ is a
parameter of the zero-order state, it does not depend upon $\ob$. 
From Krause and R\"adler (1980, p.~75) an isotropic tensor like
$R_{de}(\k)$ can be expressed in the form
\beq
R_{de}(\k) &=& (A-k^{-1}\part_k B) \delta_{de} - \lrp{ \part^2_k B-k^{-1}
\part_k B} k^{-2}k_d k_e \nn\\
&& +\ ik^{-1}\part_k C \ep_{def}k_f\; , \lab{57}
\eeq
where $A, B$, and $C$ are functions of $k$ alone.  Therefore
\be
\ep_{cde} R_{de}(-\k) = -2ik_c k^{-1} \part_k C \lab{58} \ee
is nonvanishing only for turbulence lacking reflection symmetry, and so (\ref{56})
becomes
\be
\lra{ \ep_{cde} \hz_d^{+(0)} (\k,t) \hz_e^{-(0)} (\k',t_1)} = -2i k_c k^{-1}
\partial_k C e^{-\gamma_k |t-t_1|}\;
.\lab{59}\ee

 The integration  of (\ref{55}) over
$\k'$ converts
$\k'$ to
$-\k$, so
$(\k+\k')\cdot
\x=0$.  Since $|t-t_1|=t-t_1$ in the integration over $t_1$, we find that
(\ref{55}) equals
\beq
&&- \int d\k \k\cdot \OB k_c k^{-1} \partial_k C \int^t_{-\infty}
dt_1 \exp \l\{ [i\k\cdot \OB-\la k^2-\gamma_k](t-t_1)\r\} \nn\\
&&\qquad  =  i\int d\k {i\k\cdot
\OB\over (\la k^2+\gamma_k)-i\k\cdot \OB}k_c k^{-1} \partial_k C  \; . 
\lab{60}
\eeq

We note that the second term in (\ref{51}) can be written
\be
+\tfrac12 \ep_{cde} \lra{z'^+_e (\x,t) z^{-(0)}_d (\x,t)} \lab{61}\ee
because of the $\ep_{cde}$.  Then reference to (\ref{52}) and (\ref{53}) shows that this
is the negative of (\ref{60}), with $i\leftrightarrow -i$.  Hence the
sum of the first two terms in (\ref{51}) can be written
\beq
&&\hskip-36truept i\int d\k k_c k^{-1} \partial_k C\l[ {i\k\cdot
\OB\over (\la k^2+ \gamma_k)-i\k\cdot \OB} + {i\k\cdot \OB\over (\la k^2
+\gamma _k)+i\k\cdot \OB}\r]
\nn\\
&& \hskip-8truept = -2 \int d\k k_c k^{-1} \partial_k C  \k\cdot \OB (\la
k^2+
\gamma_k) [(\la k^2+ \gamma_k)^2+(\k\cdot \OB)^2 ]^{-1}\; . 
\lab{62} \eeq

We now turn to the third term in (\ref{51}), which is
\beq
&& \hskip-12truept
-\half \int d\k \int d\k' e^{i(\k +\k')\cdot \x} (\k\cdot \OB) (\k'\cdot
\OB)\nn\\
&&  \cdot 
\int^t_{-\infty} dt_1 \exp \l\{ - [-i\k\cdot \OB+\la k^2](t-t_1) \r\}
\nn\\
&&  \cdot \int^t_{-\infty} dt_2 \exp \l\{-[i\k'\cdot \OB+\la
k'^2](t-t_2)\r\}  \lra{\ep_{cde}\hz^{+(0)}_d (\k,t_1) \hz_e^{-(0)} (\k',t_2)}\;
. \nn\\
\noalign{\vskip-8truept}
&& \lab{63}\eeq 
\vglue-16truept
\noindent According to (\ref{59}), the average is
\be
\lra{\ep_{cde}\hz^{+(0)}_d (\k,t_1) \hz^{-(0)}_e (\k',t_2)} = k_c k^{-1}
\partial_k C  \delta (\k+\k') e^{-\gamma_k |t_1-t_2|} \; , 
\lab{64}\ee so  integrating
(\ref{63}) over $\k'$ gives
\beq
&& -i \int d\k (\k\cdot \OB)^2 k_c k^{-1} \partial_k C \int^t_{-\infty}
dt_1\exp \l\{-[-i\k\cdot \OB+\la k^2] (t-t_1)\r\} \nn\\
&&\phantom{\half} \cdot \int^t_{-\infty} dt_2 \exp \l\{ -[-i\k\cdot \OB+\la
k^2](t-t_2)\r\} e^{-\gamma_k|t_1-t_2|}\; . 
\lab{65}\eeq
In this expression, the integral over $t_2$ extends from
$-\infty$ to $t_1$  with $|t_1-t_2|=t_1-t_2$, and from $t_1$ to $t$ with
$|t_1-t_2| = t_2-t_1$.  We thus find that
\beq
&&\hskip-24truept \int^t_{-\infty} dt_2 \exp \l\{ - [-i\k \cdot \OB+\la k^2]
(t_1-t_2)-\gamma_k|t_1-t_2|\r\} \nn\\
&&  = {2\gamma_k\over \gamma_k^2-(-i\k\cdot \OB+\la k^2)^2}
e^{i(\k\cdot \OB-\la k^2)(t-t_1)} + {1\over -i\k \cdot \OB+\la
k^2-\raise1pt\hbox{$\gamma_k$}} e^{-\gamma_k(t-t_1)}\; ,\nn\\
&& 
\lab{66}\eeq
so that 
\be
\int^t_{-\infty} dt_1(\quad) = \lrp{1\over i \k \cdot \B - \la k^2}
\lrp{1\over i\k\cdot \OB-\la k^2-\gamma_k}\; ,\lab{67}\ee
where (\quad) refers to the function following $dt_1$ in (\ref{65}).
This expression can be written in the equivalent form
\be
\int^t_{-\infty} dt_1 (\quad) = \gamma^{-1}_k \l[ {i\k \cdot \OB+\la
k^2\over (\k\cdot \OB)^2+(\la k^2)^2} - {i\k \cdot \OB+(\la
k^2+\gamma_k)\over (\k\cdot \OB)^2+(\la k^2+\gamma_k)^2}\r]
\;,\lab{68}
\ee so (\ref{65}) becomes
\be -i \int d\k (\k\cdot \OB)^2 k_c k^{-1} \partial_k C\gamma^{-1}_k  
 \l[ {i\k\cdot \OB+\la k^2\over (\la k^2)^2+(\k\cdot \OB)^2} -
{i\k\cdot \OB+(\la k^2+\gamma_k)\over (\la k^2+\gamma_k)^2 + (\k\cdot
\OB)^2} \r]\; . \lab{69}\ee 
We   show below that the contributions of the   terms of order
$\ob^2$ in the numerator vanish, so (\ref{69}) becomes
\be
  \int d\k (\k\cdot \OB)^3 k_c k^{-1} \partial_k C \gamma^{-1}_k \l[
{1\over (\la k^2)^2+(\k\cdot \OB)^2} - { 1\over (\la
k^2+\gamma_k)^2+(\k\cdot \OB)^2} \r] \; . \lab{70}\ee

Combining (\ref{70}) and (\ref{62}) yields the following expression for (\ref{51}):
\beq
\lra{\v\times \b}_c &=& - 2 \int d\k k_c k^{-1} \partial_k C   \l\{ {\k\cdot
\OB(\la k^2+\gamma_k)\over (\la k^2+\gamma_k)^2+(\k\cdot \OB)^2}\r.
\nn\\
&&  \l.  -
\half (\k\cdot \OB)^3 \gamma_k^{-1}
\l[ {1\over (\la k^2)^2+(\k\cdot \OB)^2} - {1\over (\la
k^2+\gamma_k)^2+(\k\cdot \OB)^2}\r]\r\} \; . \nn\\
&&\lab{71}
\eeq

To proceed, we adopt a coordinate system (1, 2, 3) such that
\be
\OB = (\ob, 0 , 0 )\lab{72}\ee
has only one component, $\ob_1=\ob$, so that
\be
\k\cdot \OB = k_1 \ob\; . \lab{73}\ee
Hence
\beq
\lra{\v\times \b}_c &=& - 2\int d\k k_c k^{-1} \part_k C \l\{ {k_1\ob(\la
k^2+\gamma_k) \over (\la k^2+\gamma_k)^2 +(k_1\ob)^2} \r.\nn\\
&&\l. - \ \half (k_1\ob)^3 \gamma_k^{-1} \l[ {1\over (\la k^2)^2 + (k_1
\ob)^2} - {1\over (\la k^2+\gamma_k)^2+(k_1\ob)^2}\r] \r\}\;.\nn\\
&&\lab{74} \eeq
Note that $\lra{\v\times \b}$ vanishes if $\ob=0$; as explained
earlier,  finite
$\ob$ is required to break rotational symmetry.  

In our coordinate
system
\be
d\k =  k^2 \sin \theta d \varphi d\theta d k \; , \lab{75}\ee
where $\theta = \angle^\k_{(1)}$, and $\varphi$ is the azimuth in the 2-3
plane. If $c = 2$ or $3$ in (\ref{74}), $k_c=k\sin \theta\sin \varphi$ or
$k\sin \theta\cos
\varphi$, and as the rest of the integrand is independent of $\varphi$, the
integral over
$\varphi$ vanishes.  We are left with $c=1$, in which case the integrand is
an even function of $k_1$.  The integral over $\varphi$ gives $2\pi$, so
\be
d\k\to - 2\pi k^2 dk d(\cos\theta) = -2\pi kdk \, dk_1\; , 
\lab{76}
\ee
where we have changed variables from $(k,\theta)$ to $(k,
k\cos\theta)=(k,k_1)$.  As the integral over $k_1$ is from $k$ to $-k$, and
the integrand is even in
$k_1$,
\be
d\k\to 4\pi k \, dk\, dk_1\; , \lab{77}\ee
with $k_1$ varying from 0 to $k$.  As claimed previously,  the terms of order
$\ob^2$ in (\ref{69}) would have contributed terms of order $k^3_1$ to the
integrand in (\ref{70}); as this is odd in $k_1$, their contribution would have
vanished when integrated from $-k$ to $k$.  Hence
\beq
\lra{\v\times \b}_1 &=& - 2 \int^\infty_{k=0} 4\pi k \, dk \int^k_{k_1=0}
dk_1\, k_1\, k^{-1} \part_k C\l\{{k_1\ob(\la k^2+\gamma_k)\over (\la
k^2+\gamma_k)^2+(k_1\ob)^2} \r.
  \nn\\
&& \l. - \half (k_1 \ob)^3 \gamma^{-1}_k \l[
{1\over (\la k^2)^2 + (k_1\ob)^2} - {1\over (\la
k^2+\gamma_k)^2+(k_1\ob)^2}\r] \r\}\; . \lab{78}
\eeq
Hence $\lra{\v\times \b}$ is parallel to $\OB$ as in the result of the
classical theory, (\ref{15}), and the dynamo coefficient
$\alpha$ is
\beq
\alpha &=& - 8\pi \int^\infty_{k=0} dk\, \part_k C \int^k_{k_1=0} dk_1 \l\{
{k^2_1(\la k^2+\gamma_k)\over (\la k^2+\gamma_k)^2+(k_1\ob)^2} \r.\nn\\
&& \l. - \half k_1^4 \ob^2 \gamma^{-1}_k \l[ {1\over (\la k^2)^2
+(k_1\ob)^2} - {1\over (\la k^2+\gamma_k)^2+(k_1\ob)^2}\r]\r\}\; . 
\lab{79}
\eeq
To carry out the integration over $k_1$ in (\ref{79}), in the first term in
braces we set
\be
\xi =  {k\ob\over \la k^2+\gamma_k}\; , \lab{80}\ee
so that
\be 
(\la k^2+\gamma_k) \int^k_{k_1=0} dk_1 {k^2_1\over (\la
k^2+\gamma_k)^2+(k_1\ob)^2} =  k^3 (\la k^2+\gamma_k)^{-1} {\xi-\tan^{-1}
\xi
\over \xi^3}\; . \lab{81} \ee 
In the first term in brackets we set
\be
\eta= {k\ob\over \la k^2}\;,\lab{82}
\ee
so that
\be 
-\half \ob^2 \gamma^{-1}_k \int^k_0 {k^4_1 dk_1\over (\la
k^2)^2+(k_1\ob)^2}   = -\half \gammak^{-1} k^3 \lrp{ {\third
\eta^3-\eta+\tan^{-1}\eta\over \eta^3}}\; . \lab{83}
 \ee 
In the second term in brackets, we again use (\ref{80}), so that 
\be
 \half \ob^2 \gammak^{-1} \int^k_0 {k_1^4 dk_1 \over (\la k^2 +
\gammak)^2+(k_1\ob)^2}  =  \half \gammak^{-1}k^3 \lrp{ {\third
\xi^3-\xi+\tan^{-1} \xi\over \xi^3}} \; .\lab{84}
\ee
Combining (\ref{81}), (\ref{83}), and (\ref{84}), we find that in (\ref{79})
\be
\int^k_{k_1=0}  dk_1 \big\{ \big\} = \half k^3 \gammak^{-1} \l[
{\gammak-\la k^2\over
\gammak+\la k^2} f(\xi)+f(\eta)\r] \;, \lab{85}\ee
where
\be
f(\xi) = {\xi-\tan^{-1} \xi\over \xi^3}\; , \lab{86}\ee
so that  (\ref{79}) becomes
\be
\alpha = -\int^\infty_{k=0} 4\pi k^2 dk\, k\part_k C\gammak^{-1} \l[
{\gammak-\la k^2\over \gammak + \la k^2} f(\xi)+f(\eta)\r] \;.\lab{87}
\ee
Since the integrand is a function of $k$ alone, $4\pi k^2 dk=d\k$ and this can
be written
\be
\alpha = -\int d\k \, k\part_k C\gammak^{-1} \l[ {\gammak-\la k^2\over
\gammak+\la k^2} f(\xi) +f(\eta)\r]\;.\lab{88}
\ee

To interpret the quantity $k \part_kC$, we multiply the
expression on the left hand side of (\ref{56}), taken for $t_1=t$, by $ik'_c$
to get
\beq
\lra{ \ep_{cde}\hz^{+(0)}_d (\k) ik'_c  z_e^{-(0)}(\k')} &=& -
\lra{\hz^{+(0)}_d (\k) [i\k'
\times \hbz^{-(0)}(\k')]_d}\nn\\
&=& -\lra{ \hbz^{+(0)} (\k)\cdot \hbw^{-(0)} (\k')}\; , \lab{89}\eeq
where 
\be
\w^{-(0)} (\x) \equiv \nabla\times \z^{-(0)} (\x)\; . \lab{90}\ee
From Leslie (1973; eq.~2.16), (\ref{89}) equals
\be
-\hat h(\k) \delta (\k +\k')\; , \lab{91}
\ee
where $\hat h(\k)$ is the transform of $h(\bxi)$, the correlation function
\be
h(\bxi) = \lra{\z^{+(0)}(\x) \cdot \w^{-(0)}(\x+\bxi)}\;.\lab{92}\ee
From (\ref{59}) with $t_1=t$, we see that (\ref{89}) is also equal to
\be 
\lra{\ep_{cde} \hz^{+(0)}_d (\k)ik'_c \hz^{-(0)}_e (\k')} 
= - ik_c (-2ik_c k^{-1}\part_k C)\delta(\k+\k') = -2k\part_k
C\delta(\k+\k')\; .\lab{93}\ee
Equating (\ref{91}) with (\ref{93}), we see that
\be
 k\part_k C=\half \hat h (\k)\;, \lab{94}\ee
so that (\ref{88}) becomes
\be
\alpha=-\half \int d\k \hat h(\k)\gamma^{-1}_k \l[ {\gamma_k-\la
k^2\over \gamma_k +\la k^2} f(\xi) +f(\eta)\r]\;, \lab{95}
\ee
which is our principal result, where $\hat h(\k), f,\xi$, and $\eta$ are
defined  by (\ref{92}), (\ref{86}), (\ref{80}), and (\ref{82}) respectively. 
 
To compare (\ref{95}) with the classical result (\ref{16}), we take the   limit
$\ob\to 0$, which according to (\ref{80}) and (\ref{82}) corresponds to $\xi\to
0,\eta\to 0$, in which case $f(\xi) = f(\eta) \to 1/3$, and the bracket in
(\ref{95}) becomes
\be
\third \lrp{ {\gammak-\la k^2\over \gammak +\la k^2} +1} = \tfrac23
{\gammak\over \gammak+\la k^2}\;, \lab{96}\ee
so  
\be
\alpha(\ob\to 0) = -\third \int d\k \hat h(\k) (\gammak+\la k^2)^{-1}\; .
\lab{97} \ee
As the  integral of the power spectrum over $\k$ is the correlation at zero
lag,
\beq
\int d\k \hat h(\k) &=& \lra{ \z^{+(0)} (\x,t)\cdot \nabla\times
\z^{-(0)}(\x,t)}
\nn\\
&=& \lra{(\v^{(0)}+\b^{(0)})\cdot \nabla\times (\v^{(0)}-\b^{(0)})} \nn\\
&=& \lra{\v^{(0)} \cdot \nabla\times \v^{(0)}} -
\lra{\b^{(0)}\cdot\nabla\times \b^{(0)}} + \lra{\nabla\cdot (\v^{(0)}\times
\b^{(0)})} \nn\\
&=& \lra{\v^{(0)}\cdot \nabla\times \v^{(0)}} - \lra{ \b^{(0)}\cdot
\nabla\times \b^{(0)}} \lab{98}
\eeq
because $\lra{\v^{(0)}\times \b^{(0)}}=0$ according to (2); here  we have
suppressed the arguments for clarity.

From (\ref{97}) we conclude that if $\la$ is small (but not too small --- see
below), and in the special case that $\gammak=\gamma_0$ is independent of
$k$,
\beq
\alpha(\ob\to 0) &=& -\third \int d\k \hat h(\k)\gamma_0^{-1}\nn\\
&=& -\third \gamma_0^{-1} \l[ \lra{\v^{(0)}(\x,t)\cdot\nabla\times
\v^{(0)}(\x,t)} \r.\nn\\
&&\quad \l. - \lra{\b^{(0)} (\x,t)\cdot\nabla\times \b^{(0)}(\x,t)}\r] \;.
\lab{99}\eeq
The first term agrees with classical result (\ref{16}) if the damping constant
$\gamma_0$ is identified with $\tau^{-1}$.    The second term
has the sign and form expected from the work of Pouquet \etal (1976), as
indicated in (\ref{17}).

Our more general result for $\ob\to 0$, (\ref{97}), differs from the classical
result in three ways:
\begin{itemize}
\item[i)] a term proportional to the current helicity is included in the form
derived by Pouquet \etal (1976);
\item[ii)] allowance is made for variation of the damping constant with
$k$; 
\item[iii)] and the effect of finite $\la$ is included.
\end{itemize}

In the opposite limit $\ob\to \infty$, $f(u) \to u^{-2}$, so the bracket in
(\ref{95}) becomes
\be 
{\gamma_k - \la k^2\over \gammak +\la k^2} {1\over u^2}+{1\over v^2}  = 
{\gammak-\la k^2\over \gammak+\la k^2} {(\gammak+\la k^2)^2\over
(k\ob)^2} + {(\la k^2)^2\over (k\ob)^2}   = \lrp{
{\gammak\over k\ob}}^2 \;,
\lab{100}\ee 
 independent of $\la$.  Thus,
\be
\alpha(\ob\to \infty) = -{1\over 2\ob^2} \int d\k \hat{h} (\k) k^{-2}
 \gammak
\;.
\lab{101}\ee
An inverse dependence on $\ob^2$ was cited by Krause and R\"adler (1980),
but with a different factor. 

 Finally, we consider the simplification
introduced when the spectral density
$C(k)$ is effectively concentrated at some wave number $k_\ast$, which if
 $k_\ast\sim 2k_0$ is  a crude representation of Figure~1 of the
numerical result of Pouquet
\etal (1976), where
$E^V_k$ peaks at $k_0 \simeq 1$, $E^M_k$ peaks at $k=3$, and the energy
beyond
$k=3$ is Alfv\'enic, so it does not contribute to $\alpha$.  Then from (\ref{80}),
(\ref{82}), and (\ref{95}),
\be
\alpha = -\half \int d\k \hat h (\k) \gamma_\ast^{-1} \l[ {\gamma_\ast - \la
k_\ast^2\over \gamma_\ast+\la k^2_\ast} f(\xi_\ast)+f(\eta_\ast)\r] \;,
\lab{102}
\ee
where $\gamma_\ast=\gamma_k(k_\ast)$,
\be \xi_\ast = {k_\ast \ob\over \gamma_\ast +\la k^2_\ast}\; , 
\lab{103}\ee 
and
\be
\eta_\ast = {k_\ast\ob\over \la k^2_\ast}\;. \lab{104}\ee
In the light of (\ref{99}), then,
\be 
\alpha  =  - \half \gamma^{-1}_\ast \l[ \lra{\v^{(0)}\cdot \nabla\times
\v^{(0)}} - \lra{\b^{(0)}\cdot \nabla\times \b^{(0)}}\r]  \l[ {\gamma_\ast - \la
k^2_\ast\over \gamma_\ast+\la k^2_\ast} f(\xi_\ast) +f(\eta_\ast)\r] \; ,
\lab{105}
\ee  so the dependence
upon $\ob$ is entirely in the multiplicative factor.  If we define a magnetic
Reynolds number by
\be
R_M = {\gamma_\ast\over \la
k_\ast^2}\;,\lab{106}
\ee
 and  a nondimensional mean magnetic field by
\be
\beta  = {\ob\over (\gamma_\ast/k_\ast)}\;, \lab{107} \ee
  we can write the bracket in (\ref{105}) in the form,
\be
F(R_M , \beta ) = {R_M-1\over R_M+1} f \lrp{\beta 
{R_M\over R_M+1}} +f(\beta R_M)\;.\lab{108}\ee
As $\beta \to 0$, this approaches
\be 
F(R_M,\obet\to 0)  =  \third \lrp{ {R_M-1\over R_M+1} +1}  
  =   \tfrac23 {\gamma_\ast \over \gamma_\ast
+\la k^2_\ast}\; , \lab{109}\ee 
which, when inserted into (\ref{105}), yields agreement with (\ref{99}), our previous
result for $\ob\to 0$.  

However, one must be careful about this procedure if $R_M$ is $\gg 1$, as
is usually the  case, for even if $\obet \ll 1$, $\obet$ may   be $\gg
R^{-1}_M$, so that the second term on the right of   (\ref{108}) is effectively 0
rather than
$1/3$.  In this case 
\be
F(R_M \to \infty, \obet\to 0) = \third\;,\lab{110}\ee
and $\alpha$ is given by $1/2$ the value in (\ref{99}).  Evidently, the classical
expression for $\alpha$ (as modified by Pouquet {\it et al.}) is off by a
factor of 2 if $R_M\gg 1$.  Figure 1 is a plot of $F(R_M,\obet)$, compared
with $F^\ast (R_M,\obet)= (1+R_M\obet^2)^{-1}$ from Cattaneo and Hughes
(1996).  As $F^\ast(R_M,\beta )$ was not proposed to be
accurate for $R_M<1$, the curves of $F^\ast$ for $R_M<1$ should be ignored. 
In qualitative agreement with $F(R_M,\beta )$, $F^\ast$ decreases as
$R_M$ increases, but of critical importance, it decreases to 0, rather than
remaining finite as $R_M\to \infty$, the case for the galactic dynamo.   

\section{Discussion}
Our general result for $\alpha$, (\ref{95}), displays significant features of the
$\alpha$-effect due to driven helical turbulence.  Our discussion of (\ref{99}),
above, makes clear that the values of $\v$ and $\b$ referred to in $\hat
h(\k)$ are $\v^{(0)}$ and $\b^{(0)}$, the values of the small-scale velocity
and magnetic fields that apply if $\OB=0$.   

As indicated in the text, it is in principle   possible to calculate $\v^{(0)}$
and
$\b^{(0)}$, and hence $\hat h(\k)$, from numerical simulations of driven
helical turbulence with $\OB=0$.  Then evaluating $\alpha$ from (\ref{95})
requires the value of $\gammak$, which can be calculated from (\ref{59}) in
terms of the statistics of the zero-order state.  One should also calculate
$\alpha$ directly from its definition using simulations with various values
of
$\ob$, and then compare the results with (\ref{95}).

Another feature of our work is that the contribution proportional to the
current helicity, $\lra{\b^{(0)}\cdot \nabla\times \b^{(0)}}$, first described
by Pouquet \etal (1976), emerges naturally from our work.  To give it
a physical interpretation, we refer to a derivation we carried out in terms
of the variables $\v^{(0)}$ and $\b^{(0)}$ to check the derivation presented 
here in terms of $\z^{\pm(0)}$.  Referring to (\ref{23}), we find that at first
order in $\OB$, the current helicity term is traceable to
$\lra{\v^{(1)}\times \b^{(0)}}$, as opposed to $\lra{\v^{(0)}\times \b^{(1)}}$,
which gives rise to the kinetic helicity term, where the velocity
perturbation $\v^{(1)}$ is due to the first-order Lorentz force
$(\nabla\times \b^{(0)}) \times \OB$ as explained by Pouquet \etal (1976;
p.~332).  

Current helicity is discussed in the Soviet and
Russian literature.  Vainshtein (1972) and Vainshtein and Zeldovich (1972)
called attention to effects of current helicity (which they referred to as
``magnetic gyrotropy").  They argued on physical grounds that as $\ob$
grows, it induces current helicity, which, because its sign is opposite to
that of $\lra{\v\cdot \nabla\times \v}$, reduces $\alpha$, causing dynamo
activity to cease at some finite value of $\ob$.  As we have explained,
Pouquet \etal (1976) showed that current helicity is important at all
values of $\ob$, and on the basis of their spectral equations, found that the
effect described by Vainshtein (1972) would be smaller and of the opposite
sign than he claimed.  Vainshtein and Kichanitov (1983) accept the
interpretation of Pouquet \etal (1976).  They also derive a differential
equation for the small-scale current helicity from the conservation of
magnetic helicity as did Kleeorin and Ruzmaikin \etal (1982); see Kleeorin
\etal (1995).  We discuss these matters further in a forthcoming paper
(Blackman and Field 1998).

The present work   relies upon $\tau\ll t_{ed}$
to evaluate the perturbations of
$\v$ and
$\b$ due to $\OB$ (although not in evaluating $\v^{(0)}$ and $\b^{(0)}$). 
This is a reasonable approximation because  the dominant
contribution to $\alpha$ comes from wavenumbers $<3k_0$, which are
relatively unaffected by magnetic  back reaction, and therefore should
follow  Pope's (1994) finding that in hydrodynamic turbulence,
$\tau<t_{ed}$.  Unfortunately, the inequality is   not as strong
as we desire.

\section{Conclusions}
We have found an analytic formula for the dynamo coefficient $\alpha$
based upon a perturbation expansion in $\ob$, the magnitude of the 
large-scale magnetic field, which is responsible for perturbing isotropic
helical turbulence in such a way as to produce a turbulent emf along
$\OB$.  Our formula gives $\alpha$ in terms of $\ob$ and the magnetic
diffusivity
$\la$, together with the spectra of   kinetic and  current helicities,
and a damping coefficient.  Both of the latter are calculable from a
simulation of incompressible isotropic helical MHD turbulence.  For small
values of
$\ob$, our results agree with the classical results as modified by Pouquet
\etal (1976).  For large
$\ob$, we find that two terms contribute to $\alpha$.  The first term  is
independent of the magnetic diffusivity $\lambda$, and is similar to the
expression proposed by  Kraichnan (1979).  The second term, which depends
on $\lambda$ in a way reminiscent of the expressions
suggested by Cattaneo and his collaborators, vanishes in the limit $\la\to
0$, as predicted by Cattaneo, {\it et al.}, but contrary to Cattaneo {\it et
al.},  the first term remains finite, and so  $\alpha$ is reduced only by a
factor of 2 for large values of
$\ob$.

\acknowledgements
We are grateful to Amitava Bhattacharjee,  B. D. G. Chandran, and Eugene
Parker for their insightful
comments, and to the Aspen Center for Physics, where part of the work
was done.  We are also grateful to Mr. Dimitri Kalaitzidakis for his
hospitality and allowing us to use his facilities while at a meeting on
Crete.  This work was supported in part by NASA Grant NAGW-931.

\appendix
\section{Energy Cascade}
Start with the Navier-Stokes equation, which describes incompressible
hydrodynamic (HD) turbulence:
\be
\part_t\v=-\v\cdot\nabla\v-\nabla p+\nu \nabla^2 \v+\f\;, 
\ee
where $p=$ pressure $\div \rho$.  Define the kinetic energy $E^V$ 
by
\be
E^V=\lra{\half v^2}\; . 
\ee
Then   the scalar product of (A1) with $\v$ shows that
\be
\part_t E^V = -\nu\lra{\omega^2}+\lra{\f\cdot \v}\;,
\ee
where
\be
\bfo = \nabla\times \v 
\ee
is the vorticity.  Hence in unforced ideal HD, $E^V$ is conserved.  The only
effect of the nonlinear term $-\v\cdot \nabla\v$ in (A.1) is to redistribute
$E^V_k$, the  kinetic energy spectrum, over $k$, keeping $E^V=\int
E^V_kdk$ constant.  Numerical simulations show that this redistribution
takes the form of a direct cascade, that is, from lower $k$ to higher $k$.

Now add a magnetic field $\B$, so the momentum equation becomes (\ref{23}):
\begin{eqnarray}
\part_t \v &=& -\v\cdot \nabla\v - \nabla p +\B\cdot \nabla\B-\nabla \half
B^2 +\nu \nabla^2 \v+\f \nonumber\\
&= &-\v \cdot \nabla \v+\B\cdot \nabla \B - \nabla P+\nu \nabla^2\v +\f\; ,
\end{eqnarray}
where
\be
P=p+\half B^2\; . 
\ee
The  equation for $E^V$ becomes
\be
\part_t E^V= \lra{\v\cdot (\B\cdot \nabla)\B} - \nu \lra{\omega^2} +
\lra{\f\cdot \v}\; ,
\ee
so $E^V$ is not conserved in unforced ideal MHD because the first
term allows magnetic energy to exchange with kinetic energy.  To account
for this, we write the induction equation 
governing $\B$:
\be
\part_t \B=-\v\cdot \nabla \B+\B\cdot \nabla \v+\nu \nabla^2\B\;, 
\ee
where we have specialized to the case of unit magnetic Prandtl number,
$\la/\nu$.  From the scalar product of  (A.8) with $\B$  we obtain an
equation for the magnetic energy
$E^M  =  \lra{\half B^2}$, \be
\part_t E^M =  \lra{\B\cdot (\B\cdot \nabla)\v}-\nu \lra{J^2}\;, 
\ee 
where
\be
{\bf J} = \nabla\times \B \; . 
\ee
$E^M$ is not conserved in unforced ideal MHD because of exchange with
$E^V$.   However, we note that in (A.10)
\begin{eqnarray}
\lra{\B\cdot (\B\cdot \nabla)\v} &=& \lra{\B\cdot \nabla (\v\cdot \B)} -
\lra{\v\cdot (\B\cdot \nabla)\B} \nonumber\\
&=& \lra{\nabla\cdot \l[ \B(\v\cdot \B)\r]} - \lra{\v\cdot (\B\cdot
\nabla)\B} \nonumber\\
&=& - \lra{\v\cdot (\B\cdot \nabla)\B} 
\end{eqnarray}
since spatial averaging converts the divergence to a vanishing
surface integral.  Hence
\be
\part_t E^M = -\lra{\v\cdot(\B\cdot \nabla)\B} - \nu \lra{J^2}\;. 
\ee
The energy-exchange term in (A12) is the negative of that in (A7), indicating
that energy gained by $E^V$ is lost by $E^M$ and {\it vice-versa}.  Hence if
$\nu=\f =0$, the sum of (A.7) and (A.13) yields
\be
\part_t E\equiv \part_t (E^V+E^M) = 0 
\ee
for the conservation of the total energy $E$ in unforced ideal MHD. 
By analogy with $E$ in HD, the  second nonlinear terms in (A.5) and (A.8)
for MHD can only redistribute $E_k$ over $k$.  Numerical studies like those
of Pouquet
\etal (1976) show that in the non helical case, the effect of the nonlinear
terms in MHD  is a direct cascade of total energy.  However, because $E^V$
and
$E^M$ are not separately conserved, it is not obvious that both $E^V$ and
$E^M$ individually cascade directly in general.  

To answer this question, we turn to the Elss\"aser variables for the
small-scale field
$\z^\pm$, which in the case $\la=\nu$ satisfy (\ref{33}):
\be
\part_t \z^\pm = -\z^\mp \cdot \nabla\z^\pm - \nabla P +\nu \nabla^2
\z^\pm +\f \pm \OB\cdot \nabla\z^\pm\;.
\ee
Define
\be
E^\pm = \lra{\half (z^\pm)^2}\;. 
\ee
Then the scalar product of  (A.15) with $\z^\pm$ yields
\begin{eqnarray}
\part_t E^\pm &=& - \lra{\z^\pm \cdot (\z^\mp\cdot
\nabla)\z^\pm}-\lra{\z^\pm \cdot \nabla P}\nonumber\\
&&+ \nu \lra{\z^\pm \cdot
\nabla^2\z^\pm} + \lra{\f\cdot \z^\pm}\pm
\lra{\z^\pm \cdot (\OB\cdot \nabla)\z^\pm}\nonumber\\
&=& - \lra{(\z^\mp \cdot \nabla)\half (z^\pm)^2} - \lra{\nabla\cdot (\z^\pm
P)} \nonumber\\
&& + \nu\lra{\z^\pm \cdot \l[ \nabla(\nabla\cdot \z^\pm)-\nabla\times
\nabla\times \z^\pm\r]}\nonumber\\
&&+ \lra{\f\cdot \z^\pm} \pm \lra{\OB\cdot \nabla\half (z^\pm)^2} \;.
\end{eqnarray}
\noindent Since $\nabla\cdot \z^\pm=0$, this gives
\begin{eqnarray}
\part_t E^\pm &=& -\lra{ \nabla\cdot \l[ \half \z^\mp
(z^\pm)^2\r]}-\lra{\nabla\cdot (\z^\pm P)}\nonumber\\
&&- \nu \lra{\z^\pm \cdot \nabla\times \bfo^\pm} +\lra{\f\cdot \z^\pm}
\pm \lra{\nabla\cdot \l[ \OB \half (z^\pm)^2\r]}\;,
\end{eqnarray}
\noindent where
\be
\bfo^\pm = \nabla\times \z^\pm = \bfo\pm \bj\;. 
\ee
Since
\begin{eqnarray}
\z^\pm\cdot \nabla\times \bfo^\pm &=& \nabla \cdot \l( \bfo^\pm \times
\z^\pm\r) +\bfo^\pm \cdot \nabla\times \z^\pm\nonumber\\
&=&\nabla\cdot (\bfo^\pm \times \z^\pm) +(\omega^\pm)^2 
\end{eqnarray}
\noindent and averages of divergences vanish,
\be
\part_t E^\pm = -\nu\lra{(\omega^\pm)}^2+\lra{\f\cdot \z^\pm}\; , 
\ee
an equation identical in form to (A.3).  We conclude that if $\la=\nu$ both
$E^+$ and
$E^-$  are  individually conserved in unforced ideal MHD, and that the effect
of nonlinear interactions is to redistribute the two  energy
spectra $E^+_k$ and $E^-_k$ individually.  Note that the presence or
absence of a mean magnetic field $\OB$ does not affect this conclusion.

That leaves the question as to whether the redistribution is a direct
cascade.  Since
\begin{eqnarray}
E^\pm &=& \lra{\half (z^\pm)^2} = \lra{\half (\v\pm \B)^2} \nonumber\\
&=& \lra{\half v^2}+\lra{\half B^2}\pm \lra{\v\cdot \B}\nonumber\\
&=& E^V +E^M \pm 2K = E\pm 2K\;, 
\end{eqnarray}
\noindent where
\be
K= \half \lra{\v\cdot \B} 
\ee
is the cross helicity (Biskamp 1993, p.~179), it follows from the
conservation of $E^+$, $E^-$, and $E$ that $K$ is also conserved in unforced
ideal MHD if $\la=\nu$.  As pointed out by
Pouquet \etal (1976), if the initial state is assumed to be statistically
invariant under $\bf b \to -\bf b$, it vanishes initially and thus remains
equal to zero,  and from (A21) we have
\be
E^+ = E^- =E\;.
\ee
As   $E$ cascades
directly,   it follows that $E^+$ and $E^-$ cascade directly if $\la=\nu$.

\section{Correlations}
\setcounter{equation}{0}
Here we show that (\ref{56}) is a reasonable representation.  Let
\be
\V = \lra{\hat{\z}^{+(0)} (\k,t)\times \z^{-(0)}(\k',t_1)}\;, 
\ee
so that the average in (\ref{55}) is $V_c$.  From (\ref{48})
\begin{eqnarray}
\V &=& (2\pi)^{-6} \int \int d\x  d\x' e^{-i(\k\cdot \x+\k'\cdot \x)}
\lra{\z^{+(0)}(\x,0)\times \z^{-(0)} (\x',\tau)} \nonumber\\
&=& (2\pi)^{-6} \int \int d\x d \br e^{-i(\k+\k')\cdot \x-i\k'\cdot \br}
\lra{\z^{+(0)} (0,0)\times \z^{-(0)} (\br,\tau)}\; , 
\end{eqnarray}
 where we have put $t_1=t+\tau$ and $\x'=\x+ \br$ and have used
stationarity and homogeneity.  Carrying out the integration over $\x$ gives
\be
\V = (2\pi)^{-3} \delta (\k +\k') \hat{\bf I} (\k',\tau) \; , 
\ee
where
\be
\hat{\bf I} (\k',\tau) = \int d \br e^{-i\k'\cdot \br} \lra{\z^{+(0)}(0,0)\times
\z^{-(0)}(\br,\tau)} \;. 
\ee
If we write
\beq
\Q (\br,\tau)& =& \lra{\v(0,0)\times \v(\br,\tau)}\; , 
\\
{\bf S} (\br,\tau) &= &\lra{\b(0,0)\times \b(\br,\tau)}\; , \\
{\bf P}_v (\br,\tau)&=&\lra{\v(0,0)\times \b(\br,\tau)}\;, \end{eqnarray}
and 
\be
{\bf P}_b(\v,\tau)= \lra{\b(0,0)\times \v(\br,\tau)}\;, 
\ee
where we have omitted superscripts for clarity, then the average in (B4) is,
from (\ref{18}) and (\ref{19}),
\begin{eqnarray}
\lra{\z^{+(0)}(0,0)\times \z^{-(0)} (\br,\tau)} &=&
\Q(\br,\tau)-{\bf S}(\br,\tau)\nonumber\\
&&- {\bf P}_v(\br,\tau) + {\bf P}_b(\br,\tau)\; , 
\end{eqnarray}
\noindent so
\be
\hat{\bf I} = \hat\Q(\k',\tau) - \hat{\bf S}(\k',\tau)-\hat{\bf P}_v
(\k',\tau)+\hat{\bf P}_b(\k',\tau)\;,
\ee
where
\be
\hat{\Q} (\k',\tau) = \int d\br e^{-i\k'\cdot \v} \Q(\br,\tau)\;, 
\ee
etc.  One of us has shown (Chou \& Fish 1998) that the terms ${\bf P}_v$ and
${\bf P}_b$ do not contribute to the final result, so we ignore them in what
follows.

Consider
\begin{eqnarray}
\Q(\br,\tau)  &=& \lra{\v(0,0)\times \v(\br,\tau)} \nonumber\\
&=& \lra{\v(-\br,-\tau)\times \v(0,0)}\nonumber\\
&=&-\lra{\v(0,0)\times \v(-\br,-\tau)}\; ,
\end{eqnarray}
\noindent
 where we have used homogeneity and stationarity  and  flipped the cross
product.  Thus
\begin{eqnarray}
\hat\Q(\k',\tau)& =& -\int d\br e^{-i\k'\cdot \br}
\lra{\v(0,0)\times
\v(-\br,-\tau)} 
\nonumber\\
& =& -\int d\br e^{i\k'\cdot {\bf p}} \lra{\v(0,0)\times
\v({\bf p},-\tau)}
\nonumber\\
& =& -\hat{\Q}(-\k',-\tau)\; .
\end{eqnarray}
\noindent 
Krause and R\"adler (1980, p.~75) show that quantities like $\Q$ are odd
functions of $\k'$, so (B13) implies that
\be
\hat\Q(\k',\tau)=\hat\Q(\k',-\tau)\; . 
\ee
Since ${\bf S}(\br,\tau)$ has a similar structure,
\be
\hat{\bf S}(\k',\tau) = \hat{\bf S}(\k',-\tau)\; , 
\ee
so that $\hat{\bf I}(\k',\tau)$ is an even function of $\tau$, or, equivalently,
a function of $|\tau|$, which we shall denote $(2\pi)^3  {\bf
R}(\k',|\tau|)$.  Hence from (B3)
\be
V_c= \delta (\k+\k')  R_c(\k',|\tau|) \;.
\ee
Because $R$ is composed of cross products, we introduce a tensor
$R_{de}(\k',|\tau|)$ such that
\be
R_c (\k',|\tau|) = \epsilon_{cde}R_{de}(\k',|\tau|)\; , 
\ee
so that
\be
V_c = \epsilon_{cde}\delta (\k+\k')R_{de}(\k',|\tau|)\; . 
\ee

The last step is to assume that $R_{de} (\k',|\tau|)$ is a function of $\k'$
times a function of $|\tau|$ (which may depend upon $k=|\k|=|\k'|$):
\be
R_{de} (\k',|\tau|) = R_{de} (\k')f_k(|\tau|)\; , 
\ee
where by taking $f_k(0)=1$, we assure that $R_{de}(\k')$ represents the
maximum value of the correlation.  As a natural choice for $f_k(|\tau|)$ we
take
\be
f_k(|\tau|) = e^{-\gamma_k|\tau|}\;,
\ee
where $\gamma_k$ is a $k$-dependent inverse correlation time, in the
spirit of the eddy-damping approximation of Pouquet et al.\ (1976) and
Kraichnan (1979).  Then
\be
V_c = \epsilon_{cde} \delta(\k+\k')R_{de}(\k')e^{-\gamma_k|\tau|}\; ,
\ee
which is (\ref{56}).

\clearpage

\figcaption{Three functions of the magnetic Reynolds number
$R_M$ and the normalized magnetic field $\beta  =
\overline{B}/v_0$ are plotted and compared.  $F(R_M,\beta)$, defined in
(110) and represented
by solid lines, is our result for the dependence of $\alpha$ on $R_M$ for
$\beta =10^{-5}$, $3\times 10^{-3}$, and 1.
$F(R_M,\beta \to 0)$, defined in (109) and represented by the
circles, is our result for $\beta \to 0$.
$F^\ast(R_M,\beta )=(1+R_M\obet^2)^{-1}$ is the dependence of
$\alpha$ suggested by Cattaneo \& Hughes (1996); as it was not intended to
be accurate for $R_M<1$, points for $R_M<1$ should be ignored.}
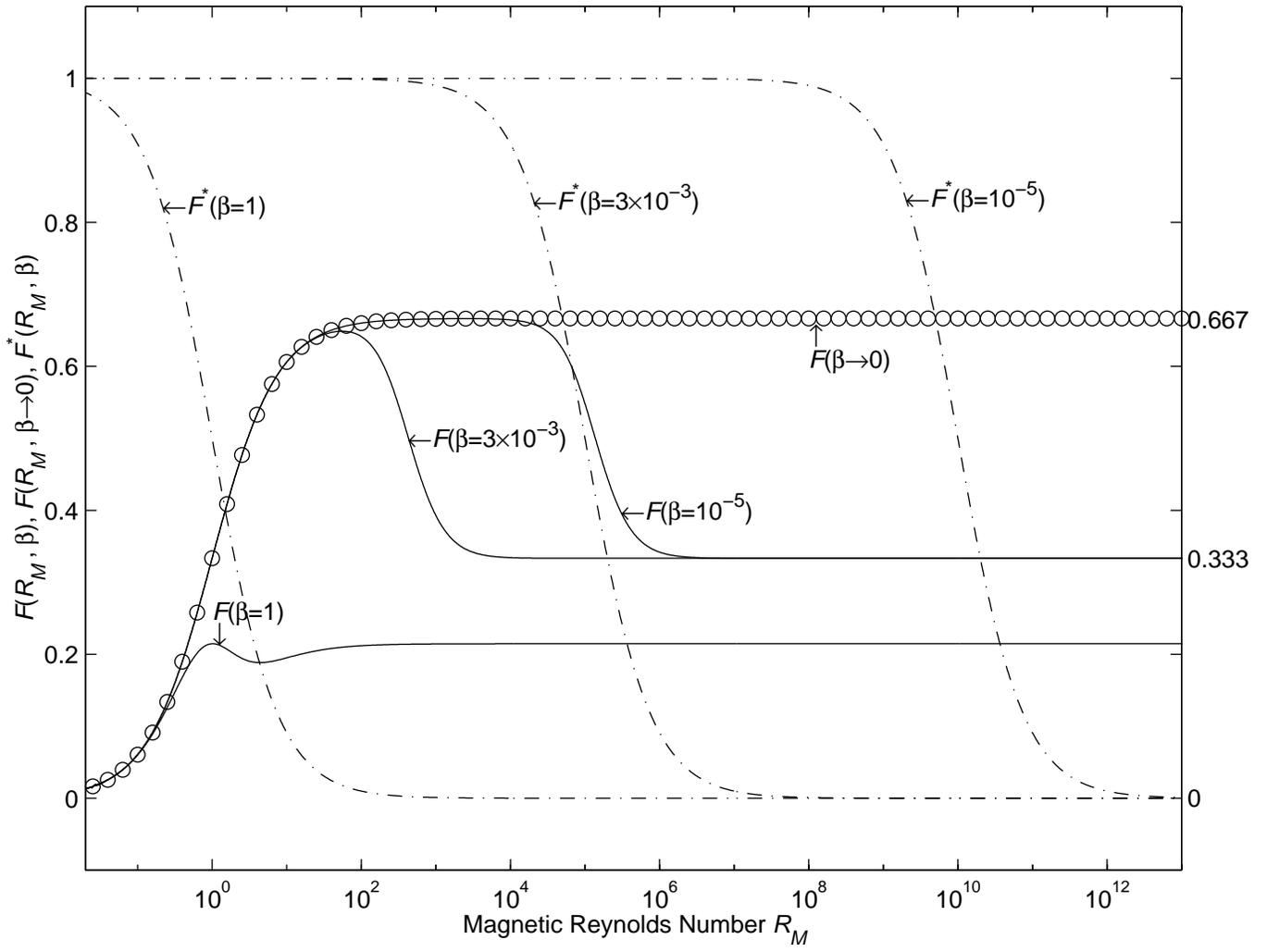
\end{document}